\documentclass[epj]{svjour}

\usepackage[ngerman, english]{babel}   
\usepackage[utf8]{inputenc}
\usepackage{lmodern}
\usepackage{amssymb}
\usepackage{amsmath}
\usepackage{graphicx}
\usepackage{diagbox} % for making diagonally-split cells in tables
\usepackage{hhline}  % for interrupting horizontal and vertical lines in tables

\usepackage{soul}

\usepackage{xr} % for equation and figure numbers
\usepackage{cite} % needed for multiple citations \cite{A,B,C}
  
\usepackage[usenames,dvipsnames]{xcolor} % text colors
\usepackage[colorlinks, citecolor={blue!50!black}, urlcolor={blue!50!black}, linkcolor={red!50!black},hypertexnames=false]{hyperref}
\usepackage[figure]{hypcap}

\newcommand{\up}{\uparrow}
\newcommand{\dn}{\downarrow}
\newcommand{\kv}{\ensuremath{\mathbf{k}}}
\newcommand{\qv}{\ensuremath{\mathbf{q}}}

\newcommand{\ch}{\ensuremath{\text{ch}}}
\newcommand{\sz}{\ensuremath{\text{sp}}}
\newcommand{\sing}{\ensuremath{\text{s}}}

\newcommand{\pp}{\ensuremath{{pp}}}
\newcommand{\ph}{\ensuremath{{ph}}}
\newcommand{\phv}{\ensuremath{\overline{ph}}}

\usepackage{subfigure}   % subfigure
\usepackage{tikz}             % draw within latex

\usetikzlibrary{positioning}
\usetikzlibrary{arrows.meta}
\usetikzlibrary{decorations.markings}
\usetikzlibrary{decorations.pathmorphing}
\usetikzlibrary{decorations.pathreplacing}
\usetikzlibrary{calc}
   
\tikzset{
   linePlain/.style={draw=black, thick},
   lineBare/.style={draw=gray, thick},
   lineWithArrowEnd/.style={draw=black, thick, postaction={decorate},decoration={markings,mark=at position 1. with {\arrow[scale=1.2]{latex}}}},
   lineBareWithArrowEnd/.style={draw=gray, thick, postaction={decorate},decoration={markings,mark=at position 1. with {\arrow[scale=1.2]{latex}}}},
   lineWithArrowCenter/.style={draw=black, thick, postaction={decorate},decoration={markings,mark=at position .6 with {\arrow[scale=1.2]{latex}}}},
   lineWithArrowCenterCenter/.style={draw=black, thick, postaction={decorate},decoration={markings,mark=at position .5 with {\arrow[scale=1.2]{latex}}}},
   lineBareWithArrowCenter/.style={draw=gray, thick, postaction={decorate},decoration={markings,mark=at position .6 with {\arrow[scale=1.2]{latex}}}},
   lineWithArrowCenterEnd/.style={draw=black, thick, postaction={decorate},decoration={markings,mark=at position .85 with {\arrow[scale=1.2]{latex}}}},
   lineBareWithArrowCenterEnd/.style={draw=gray, thick, postaction={decorate},decoration={markings,mark=at position .85 with {\arrow[scale=1.2]{latex}}}},
   lineWithArrowCenterStart/.style={draw=black, thick, postaction={decorate},decoration={markings,mark=at position .35 with {\arrow[scale=1.2]{latex}}}},
   lineWithArrowInline/.style={draw=black, semithick, postaction={decorate},decoration={markings,mark=at position .7 with {\arrow[scale=1.2]{latex}}}},
   vertex/.style={draw, shape=circle, fill=black, minimum size=1.1mm, inner sep=0mm, outer sep=0mm},
   bosonLine/.style={draw=black, thick, decorate, decoration={snake, segment length=2mm, amplitude=0.6mm}},
}
\newcommand{\tikzm}[2]{
   \tikz[baseline=-0.65ex]{#2}
}

\newcommand{\bosonfull}[4]{
   \draw[bosonLine, very thick] (#1,#2) -- (#3,#4);
}

\newcommand{\arrowslefthalf}[2]{
   \def\shift{0.3};
   \coordinate (center) at (#1,#2);
   \draw[lineWithArrowCenterEnd] (center)  -- ($(center) + (-\shift,-\shift)$);
   \draw[lineWithArrowCenterEnd] ($(center)     + (-\shift,+\shift)$) -- (center);
}
\newcommand{\arrowsrighthalf}[2]{
   \def\shift{0.3};
   \coordinate (center) at (#1,#2);
   \draw[lineWithArrowCenterEnd] ($(center) + (+\shift,-\shift)$) -- (center);
   \draw[lineWithArrowCenterEnd] (center)    -- ($(center)   + (+\shift,+\shift)$);
}

\newcommand{\arrowslefthalffull}[3]{
   \def\shift{0.3};
   \def\shiftbox{0.3*#3};
   \coordinate (center) at (#1,#2);
   \coordinate (bottomleft)  at ($(center) + (-\shiftbox,-\shiftbox)$);
   \coordinate (topleft)     at ($(center) + (-\shiftbox,+\shiftbox)$);
   \draw[lineWithArrowCenterEnd] (bottomleft)  -- ($(bottomleft) + (-\shift,-\shift)$);
   \draw[lineWithArrowCenterEnd] ($(topleft)     + (-\shift,+\shift)$) -- (topleft);
}
\newcommand{\arrowsrighthalffull}[3]{
   \def\shift{0.3};
   \def\shiftbox{0.3*#3};
   \coordinate (center) at (#1,#2);
   \coordinate (bottomright) at ($(center) + (+\shiftbox,-\shiftbox)$);
   \coordinate (topright)    at ($(center) + (+\shiftbox,+\shiftbox)$);
   \draw[lineWithArrowCenterEnd] ($(bottomright) + (+\shift,-\shift)$) -- (bottomright);
   \draw[lineWithArrowCenterEnd] (topright)    -- ($(topright)   + (+\shift,+\shift)$);
}
\newcommand{\arrowslowerhalffull}[3]{
   \def\shift{0.3};
   \def\shiftbox{0.3*#3};
   \coordinate (center) at (#1,#2);
   \coordinate (bottomleft)  at ($(center) + (-\shiftbox,-\shiftbox)$);
   \coordinate (bottomright) at ($(center) + (+\shiftbox,-\shiftbox)$);
   \draw[lineWithArrowCenterEnd] (bottomleft)  -- ($(bottomleft) + (-\shift,-\shift)$);
   \draw[lineWithArrowCenterEnd] ($(bottomright) + (+\shift,-\shift)$) -- (bottomright);
}
\newcommand{\arrowsupperhalffull}[3]{
   \def\shift{0.3};
   \def\shiftbox{0.3*#3};
   \coordinate (center) at (#1,#2);
   \coordinate (topleft)     at ($(center) + (-\shiftbox,+\shiftbox)$);
   \coordinate (topright)    at ($(center) + (+\shiftbox,+\shiftbox)$);
   \draw[lineWithArrowCenterEnd] ($(topleft)     + (-\shift,+\shift)$) -- (topleft);
   \draw[lineWithArrowCenterEnd] (topright)    -- ($(topright)   + (+\shift,+\shift)$);
}

\newcommand{\barevertexwithlegs}[2]{
   \fill (#1,#2) circle (2pt) coordinate (center);
   \arrowslefthalf{#1}{#2}
   \arrowsrighthalf{#1}{#2}
}

\newcommand{\fullvertexwithlegs}[4]{
   \def\shift{0.3};   
   \def\shiftbox{0.3*#4};
   \coordinate (center) at (#2,#3);
   \coordinate (bottomleft)  at ($(center) + (-\shiftbox,-\shiftbox)$);
   \coordinate (topleft)     at ($(center) + (-\shiftbox,+\shiftbox)$);
   \coordinate (bottomright) at ($(center) + (+\shiftbox,-\shiftbox)$);
   \coordinate (topright)    at ($(center) + (+\shiftbox,+\shiftbox)$);
   \draw[linePlain, fill=verylightgray] (bottomleft) rectangle (topright);
   \node at (center) {#1};
   \draw[lineWithArrowCenterEnd] (bottomleft)  -- ($(bottomleft) + (-\shift,-\shift)$);
   \draw[lineWithArrowCenterEnd] ($(topleft)     + (-\shift,+\shift)$) -- (topleft);
   \draw[lineWithArrowCenterEnd] ($(bottomright) + (+\shift,-\shift)$) -- (bottomright);
   \draw[lineWithArrowCenterEnd] (topright)    -- ($(topright)   + (+\shift,+\shift)$);
}

\newcommand{\threepointvertexleft}[4]{
   \draw[linePlain, fill=verylightgray] (#2,#3+0.4*#4) -- (#2+0.6*#4,#3) -- (#2,#3-0.4*#4) -- (#2,#3+0.4*#4);
   \node at (#2+0.2*#4,#3) {#1};
}

\newcommand{\threepointvertexleftarrows}[4]{
   \threepointvertexleft{#1}{#2}{#3}{#4}
   \arrowslefthalffull{#2+0.4*#4}{#3}{4./3.*#4}
}

\newcommand{\threepointvertexright}[4]{
   \draw[linePlain, fill=verylightgray] (#2,#3) -- (#2+0.6*#4,#3+0.4*#4) -- (#2+0.6*#4,#3-0.4*#4) -- (#2,#3);
   \node at (#2+0.4*#4,#3) {#1};
}

\newcommand{\threepointvertexrightarrows}[4]{
   \threepointvertexright{#1}{#2}{#3}{#4}
   \arrowsrighthalffull{#2+0.2*#4}{#3}{4./3.*#4}
}

\newcommand{\threepointvertexupper}[4]{
   \draw[linePlain, fill=verylightgray] (#2-0.4*#4,#3+0.3*#4) -- (#2+0.4*#4,#3+0.3*#4) -- (#2,#3-0.3*#4) -- (#2-0.4*#4,#3+0.3*#4);
   \node at (#2,#3+0.1*#4) {#1};
}

\newcommand{\threepointvertexupperarrows}[4]{
   \threepointvertexupper{#1}{#2}{#3}{#4}
   \arrowsupperhalffull{#2}{#3-0.1*#4}{4./3.*#4}
}

\newcommand{\threepointvertexlower}[4]{
   \draw[linePlain, fill=verylightgray] (#2-0.4*#4,#3-0.3*#4) -- (#2+0.4*#4,#3-0.3*#4) -- (#2,#3+0.3*#4) -- (#2-0.4*#4,#3-0.3*#4);
   \node at (#2,#3-0.1*#4) {#1};
}

\newcommand{\threepointvertexlowerarrows}[4]{
   \threepointvertexlower{#1}{#2}{#3}{#4}
   \arrowslowerhalffull{#2}{#3+0.1*#4}{4./3.*#4}
}

\usepackage{tabularx}                    % tables with manually set width
\usepackage{multirow, bigdelim}  % combine multiple rows in table, big delimiters
\usepackage{longtable}
\usepackage{makecell}                   % multi-line cells
\usepackage{cellspace}

\usepackage{mdwlist}      % list handling

\usepackage{verbatim}       % environment for interpreting all symbols as text, in tt font
\usepackage{nicefrac}        % fractions with diagonal slash
\usepackage[vcentermath]{youngtab} % Young tableaux
\usepackage{cancel}

\usepackage{etoolbox} % if-conditions in newcommand

\newcolumntype{L}[1]{>{\raggedright\arraybackslash}p{#1}} % linksbündig mit Breitenangabe
\newcolumntype{C}[1]{>{\centering\arraybackslash}p{#1}} % zentriert mit Breitenangabe
\newcolumntype{R}[1]{>{\raggedleft\arraybackslash}p{#1}} % rechtsbündig mit Breitenangabe
\definecolor{schwarz}{RGB}{0,0,0}
\definecolor{braun}{RGB}{102,51,0}
\definecolor{blau}{RGB}{0,84,159}
\definecolor{tiefblau}{RGB}{0,0,255}
\definecolor{maigruen}{RGB}{189,205,0}
\definecolor{rot}{RGB}{204,7,30}
\definecolor{tiefrot}{RGB}{255,0,0}
\definecolor{bordeaux}{RGB}{161,16,53}
\definecolor{violett}{RGB}{97,33,88}
\definecolor{lila}{RGB}{122,111,172}
\definecolor{tieflila}{RGB}{204,0,204}
\definecolor{magenta}{RGB}{255,0,255}
\definecolor{orange}{RGB}{255,100,0}    % {255,140,0}
\definecolor{gelb}{RGB}{246,168,0}       % {247,166,60}
\definecolor{gruen}{RGB}{87,171,39}      % {85,170,67}
\definecolor{petrol}{RGB}{0,97,101}
\definecolor{rot2}{RGB}{205,2,37}        %%%%
\definecolor{blau2}{RGB}{0,86,153}      %%%%
\definecolor{darkpetrol}{RGB}{0,73,76}

\definecolor{verylightgray}{RGB}{240,240,240}

\definecolor{darkgreen}{rgb}{0,0.5,0}
\definecolor{purple}{rgb}{0.6,0,0.5}
\definecolor{orange}{rgb}{1,0.5,0}
\definecolor{darkred}{rgb}{.7,0,0}
\definecolor{darkblue}{rgb}{0,0,.3}
\definecolor{blue}{rgb}{0,0,1}
\definecolor{grey}{rgb}{.6,.6,.6}
\definecolor{dimgreen}{rgb}{0.2,0.6,0.1}
\hypersetup{colorlinks,linkcolor=blue,urlcolor=blue,citecolor=blue}

\usepackage[normalem]{ulem}

\newcommand{\al}[6]{
	^{
	\ifcase #5 %0
		\alpha_#1 \alpha_#2 
	\or %1
		\alpha_#1^\prime \alpha_#2^\prime 
	\or %2
		#1 #2
	\or %3
		\bar{\alpha}_#1 \alpha_#2
	\or %4
		\alpha_#1 \bar{\alpha}_#2
	\or %5
		\bar{\alpha}_#1 \bar{\alpha}_#2
	\or %6
		\bar{\alpha}_#1^\prime \alpha_#2^\prime
	\or %7
		\alpha_#1^\prime \bar{\alpha}_#2^\prime
	\or %8
		\bar{\alpha}_#1^\prime \bar{\alpha}_#2^\prime
	\else
		0
	\fi
	|
	\ifcase #6 %0
		\alpha_#3 \alpha_#4
	\or %1
		\alpha_#3^\prime \alpha_#4^\prime
	\or %2
		#3 #4
	\or %3
		\bar{\alpha}_#3 \alpha_#4
	\or %4
		\alpha_#3 \bar{\alpha}_#4
	\or %5
		\bar{\alpha}_#3 \bar{\alpha}_#4
	\or %6
		\bar{\alpha}_#3^\prime \alpha_#4^\prime
	\or %7
		\alpha_#3^\prime \bar{\alpha}_#4^\prime
	\or %8
		\bar{\alpha}_#3^\prime \bar{\alpha}_#4^\prime
	\else
		0
	\fi
	}
}
\newcommand{\sigm}[6]{
	_{
	\ifcase #5
		\sigma_#1 \sigma_#2 
	\or
		\sigma_#1^\prime \sigma_#2^\prime 
	\or
		#1 #2
	\else
		0
	\fi
	|
	\ifcase #6
		\sigma_#3 \sigma_#4
	\or
		\sigma_#3^\prime \sigma_#4^\prime
	\or
		#3 #4
	\else
		0
	\fi
	}
}
\newcommand{\sig}[2]{
	{
	\ifcase #1
		\sigma \sigma
	\or
		\sigma \bar{\sigma}
	\or
		\bar{\sigma} \sigma
	\else
		0
	\fi
	|
	\ifcase #2
		\sigma \sigma
	\or
		\sigma \bar{\sigma}
	\or
		\bar{\sigma} \sigma
	\else
		0
	\fi
	}
}
\newcommand{\q}[6]{
	\ifcase #5
		q_#1 q_#2 
	\or
		q_#1^\prime q_#2^\prime 
	\else
		0
	\fi
	|
	\ifcase #6
		q_#3 q_#4
	\or
		q_#3^\prime q_#4^\prime
	\else
		0
	\fi
}
\newcommand{\om}[6]{
	\ifcase #5
		\nu_#1 \nu_#2 
	\or
		\nu_#1^\prime \nu_#2^\prime 
	\else
		0
	\fi
	|
	\ifcase #6
		\nu_#3 \nu_#4
	\or
		\nu_#3^\prime \nu_#4^\prime
	\else
		0
	\fi
}

\usepackage[normalem]{ulem} % to use strikethrough command \sout{}
\definecolor{darkgreen}{rgb}{0,0.5,0}
\definecolor{purple}{rgb}{0.6,0,0.5}
\definecolor{orange}{rgb}{1,0.5,0}
\definecolor{darkred}{rgb}{.7,0,0}
\definecolor{darkblue}{rgb}{0,0,.6}
\definecolor{grey}{rgb}{.6,.6,.6}
\definecolor{dimgreen}{rgb}{0.2,0.6,0.1}

    \tikzset{middlearrow/.style={
                decoration={markings,
                            mark= at position 0.65 with {\arrow{#1}} ,
                                    },
                                            postaction={decorate}
                                                }
                                                }

\begin{document}
	\sloppy % needed so the text does not go over the right margin

\title{The plain and simple parquet approximation: single- and multi-boson exchange in the two-dimensional Hubbard model}
\titlerunning{The plain and simple parquet approximation}
\authorrunning{F. Krien and A. Kauch}

\author{
	Friedrich Krien%\inst{1}
	\and
	Anna Kauch
	%\inst{1}
}

\institute{
	Institute for Solid State Physics, TU Wien, 1040 Vienna, Austria
}

\date{\today}
%\PACS{}

\abstract{
    The parquet approach to vertex corrections is unbiased but computationally demanding.
    Most applications are therefore restricted to small cluster sizes or rely on various simplifying approximations.
    We have recently shown that the bosonization of the parquet diagrams provides interpretative
    and algorithmic advantages over the original purely fermionic formulation.
    Here we present first results of the numerical implementation of this method by applying it to the half-filled Hubbard model on the square lattice at weak coupling. The improved algorithmic performance allows us to evaluate the parquet approximation for a $16\times16$ {lattice}, retaining the full momentum and frequency structure of the various vertex functions. We discuss their symmetries and consider parametrizations of
    their momentum dependence using the truncated unity approximation.
}

\maketitle

\section{Introduction \label{sec:Introduction}}
Methods of quantum field theory represent a cornerstone of many-body physics.
In their most general form they require the computation of multi-point correlation functions,
whose dependence on several momentum and frequency labels lies beyond any practical implementation in most cases.
An elegant formalism for the derivation of computationally feasible approximations for the electronic self-energy was introduced by Hedin~\cite{Hedin65}, who expressed the latter in terms of the Green's function ($G$), the screened interaction ($W$),
and a vertex correction ($\gamma$). The simplest, so-called $GW$ approximation already includes the feedback of collective excitations on fermions and has become a standard tool of electronic structure theory, see, for example, Ref.~\cite{Kutepov16} and references therein.

It is hard to go beyond the $GW$ approximation, although it is desirable in cases of gross quantitative discrepancies
to experiment~\cite{Kutepov16} or, for example, at strong coupling where vertex corrections may alter the interaction between fermions and bosons qualitatively~\cite{Krien21}. However, it is not a trivial task to even define proper strategies to extend the $GW$ approximation:
as usual, derivability from a potential leads to approximations that respect conservation laws~\cite{Almbladh99}; on the other hand 
one may also prefer, for example, a positive semi-definite real-axis spectrum as a stringent criterion~\cite{Leeuwen14,Leeuwen16}; or aim at including strong correlation effects~\cite{Biermann03,Ayral15}.
Here, on yet a different note, we are interested in an unbiased approach to the vertex correction $\gamma$ as it is provided, for example, by the parquet approach~\cite{Diatlov57,Dominicis64-2} or by the functional renormalization group (fRG,~\cite{Metzner12,Dupuis21}), which respect the crossing symmetry of two-particle correlation functions.

Following this path, we recently introduced a variation of Hedin's equations
which is equivalent to the parquet approach~\cite{Krien21-2};
or, vice versa, one may say that the parquet approach was recast exactly into the $GW\gamma$ form.
As such, it requires as an input the fully irreducible vertex $\Lambda$ of the parquet formalism,
where fully irreducible implies that it can not be cut
into two parts by removing two Green's function lines ($GG$-irreducible,~\cite{Rohringer12}).
The quantities that appear in Hedin's equations are, however, 
irreducible with respect to the bare interaction ($U$-irreducible,~\cite{Krien21-2}).
Therefore, the reformulated parquet equations actually use $\tilde{\Lambda}=\Lambda-U$ as a fundamental building block,
where $U$ is the Hubbard interaction. In the application presented in this work we consider the parquet approximation,
where $\tilde{\Lambda}$ vanishes, leading nevertheless to a highly nontrivial approximation for the Hubbard model.

\begin{figure}
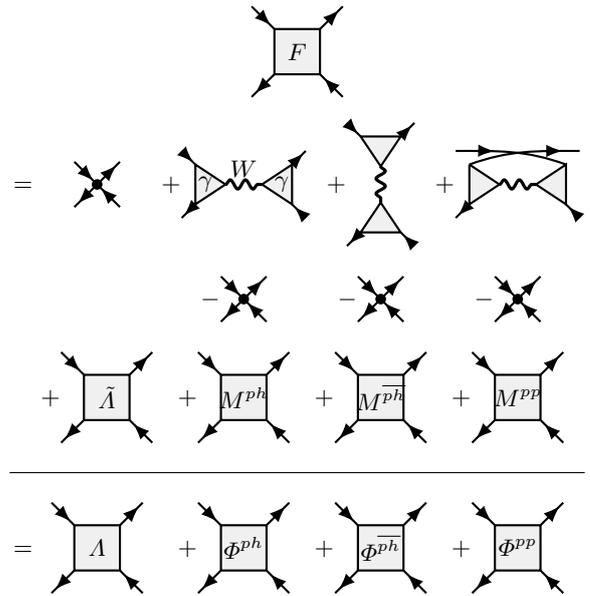

	\begin{center}
		\tikzm{Vertex_parametrization-channeldec-PE}{
			\fullvertexwithlegs{$F$}{0}{0}{1}
		}
    \end{center}
	\[\def\arraystretch{1}
	\begin{array}{ccccc}
		=&
		\tikzm{SBE_decomposition_irr}{
            \barevertexwithlegs{0}{0}
		}
		&+
			\tikzm{SBE_decomposition_a}{
			    \begin{scope}[scale=0.6]
				\threepointvertexleftarrows{\small$\gamma$}{0}{0}{1.1}
				\bosonfull{0.66}{0}{1.46}{0}
				\node at (1.06,0.35) {\small$W$};
				\threepointvertexrightarrows{\small$\gamma$}{1.46}{0}{1.1}
				\end{scope}
			}
		&+
			\tikzm{SBE_decomposition_t}{
			\begin{scope}[scale=0.6]
				\threepointvertexupperarrows{}{0}{0.73}{1.1}
				\bosonfull{0}{0.4}{0}{-0.4}
				\threepointvertexlowerarrows{}{0}{-0.73}{1.1}
			\end{scope}
			}
		&+
			\tikzm{SBE_decomposition_p}{
			\begin{scope}[scale=0.6]
				\threepointvertexleft{}{0}{0}{1.1}
				\bosonfull{0.66}{0}{1.46}{0}
				\node at (1.06,0.74) {};
				\threepointvertexright{}{1.46}{0}{1.1}
				\draw[lineWithArrowCenterEnd] (0,-0.44) -- (-0.3,-0.74);
				\draw[lineWithArrowCenterEnd] (2.42,-0.74) -- (2.12,-0.44);
				\draw[lineWithArrowCenterEnd] (0,0.44) to [out=20, in=180] (2.42,0.74);
				\draw[lineWithArrowCenterStart] (-0.3,0.74) to [out=0, in=160] (2.12,0.44);
			\end{scope}
			}\\&&&&\\
		&
		&-
			\tikzm{SBE_decomposition_a}{
                \barevertexwithlegs{0}{0}
			}
		&-
			\tikzm{SBE_decomposition_p}{
				\barevertexwithlegs{0}{0}
			}
		&-
			\tikzm{SBE_decomposition_t}{
				\barevertexwithlegs{0}{0}
			}\\&&&&\\
	    &
	    +
		\tikzm{Vertex_parametrization-channeldec-PE_R}{
			\fullvertexwithlegs{$\tilde{\Lambda}$}{0}{0}{1}
		}
		&+
		\tikzm{Vertex_parametrization-channeldec-PE_a}{
			\fullvertexwithlegs{$M^{\ph}$}{0}{0}{1}
		}
		&+
		\tikzm{Vertex_parametrization-channeldec-PE_p}{
			\fullvertexwithlegs{\footnotesize$M^{\phv}$}{0}{0}{1}
		}
		&+
		\tikzm{Vertex_parametrization-channeldec-PE_t}{
			\fullvertexwithlegs{\footnotesize$M^{\pp}$}{0}{0}{1}
		}\\&&&&\\\hline&&&&\\
    =&
		\tikzm{Vertex_parametrization-channeldec-PE_R}{
			\fullvertexwithlegs{$\Lambda$}{0}{0}{1}
		}
		&+
		\tikzm{Vertex_parametrization-channeldec-PE_a}{
			\fullvertexwithlegs{$\Phi^{\ph}$}{0}{0}{1}
		}
		&+
		\tikzm{Vertex_parametrization-channeldec-PE_p}{
			\fullvertexwithlegs{$\Phi^{\phv}$}{0}{0}{1}
		}
		&+
		\tikzm{Vertex_parametrization-channeldec-PE_t}{
			\fullvertexwithlegs{$\Phi^{\pp}$}{0}{0}{1}
		}\\
	\end{array}
	\]
\caption{Traditional and bosonized parquet decomposition, drawn below and above the horizontal line, respectively.
The Hedin vertex (triangles) and the screened interaction (wiggly lines)
are bold diagrammatic building blocks not used in the traditional formalism.
Arrows indicate attached Green's function legs, dots the bare interaction.
Notice that equality holds for each column separately;
here we focus on the horizontal particle-hole channel [second column, cf. Eq.~\eqref{eq:phi}].}
\label{fig:sbe_parquet}
\end{figure}

To put the unification of Hedin's formalism with the parquet approach into perspective,
we recall that a key technique of quantum field theory is the {\sl boldification} of Feynman diagrams:
summarily denoting a partial series of diagrams by an effective quantity, as for example the self-energy,
reduces the number of Feynman diagrams that need to be evaluated,
at the expense of keeping track of the effective quantity which has to be computed self-consistently.
In Hedin's original approach further diagrams are summarized in the screened interaction $W$ and in the polarization,
representing, respectively, a boson and a bosonic self-energy.
The corresponding reduction in the number of Feynman diagrams is concisely put on display in Ref.~\cite{Molinari06},
where it is also noted that keeping track of yet another quantity,
the $U$-irreducible Hedin vertex $\gamma$ which mediates a Yukawa-like coupling between fermions and bosons,
can be used to boldify diagrams even further.
In this spirit, the key theoretical step taken in Ref.~\cite{Krien21-2} is to boldify a subset of diagrams arising from the Bethe-Salpeter equations, which are of a simple structure.
Namely, the $U$-reducible diagrams, coined {\sl single-boson exchange (SBE)} in Ref.~\cite{Krien19-2},
are representable in terms of the bold objects $\gamma$ and $W$ (see Fig.~\ref{fig:sbe_parquet}).
The remaining $U$-irreducible diagrams, to which we refer as multi-boson exchange ($M$),
do not permit a representation in terms of $\gamma, W$ alone, but instead capture repeated exchange of bosons.
The resulting picture of bosons mediating effective interactions~\cite{Bonetti22}
is physically appealing and remains valid even at strong coupling~\cite{Harkov21,Krien21}.

Further, it is plausible that fewer Feynman diagrams correspond to reduced computational cost in practical applications.
Indeed, using the bosonized parquet approach, we are in a position to evaluate
the parquet approximation for the Hubbard model on a $16\times16$ lattice,
which is, to our knowledge, the hitherto largest cluster size reached before any approximate parametrization
of the momentum-dependent vertex functions as, for example, the truncated unity (TU) approximation~\cite{Husemann09,Wang12,Lichtenstein17,Eckhardt20}. We note in passing that the performance may be also improved
through a nonlocal formulation of the parquet approach~\cite{Krien20,Astretsov19}.
{But we refrain from applying any further approximations} (besides the parquet approximation itself).
Therefore, the computational cost is reduced here only by the asymptotic decay of the
vertex functions $M$ after the SBE diagrams are treated separately,
because the latter determine the parquet vertices asymptotically~\cite{Wentzell20}.
As a result, frequency summations involving the vertex functions $M$ decay by one power
faster compared to diagrams arising in the traditional parquet approach and, hence,
the number of Matsubara frequencies can be reduced and the momentum grid refined.
We thus arrive at the full-fledged parquet approximation for the Hubbard model,
as envisioned in the seminal papers~\cite{Diatlov57,Dominicis64-2}, progressing further along
the path of pioneering applications to the Anderson impurity model~\cite{Chen92} and small Hubbard clusters~\cite{Yang09, Tam13}.

Recently, the parquet approach has also been unified with the multi-loop functional renormalization group (mfRG,~\cite{Kugler18-3}).
By extension the latter can be recast in terms of boson exchange as well,
a corresponding theory is presented in Ref.~\cite{Walter22}.
Further efforts aim at unbiased extensions of the dynamical mean-field theory (DMFT,~\cite{Georges96})
in order to reach the strong-coupling regime~\cite{Toschi07,Taranto14,Rohringer18,Krien20}.
Implementation details of the different methods vary widely and often additional approximations need to be applied.
Therefore, we put here a spotlight on the plain parquet approximation as a (comparatively) simple reference case,
which nevertheless provides a quantitative description of the Hubbard model
in the weak-coupling limit~\cite{Hille20,Schaefer21}.
The aim of this paper is therefore twofold:
On one hand, we discuss the qualitative behavior of various correlation functions, evaluated within
the parquet approximation for the half-filled square lattice at weak coupling.
On the other hand, with the full momentum dependence of the vertex functions readily available,
we put two important tools to the test, namely,
the TU approximation~\cite{Eckhardt20} and the vertex asymptotics~\cite{Wentzell20}.
In the latter case our presentation extends to nonlocal correlations the investigation of Ref.~\cite{Harkov21},
which compared the SBE diagrams to the vertex asymptotics for the Anderson impurity model.

The paper is structured as follows. 
We recollect definitions of the bosonized parquet formalism in Sec.~\ref{sec:parquet}.
The screened interaction and Yukawa couplings are presented in Sec.~\ref{sec:wq},
various four-point vertex functions are examined in Sec.~\ref{sec:sbembe}.
The convergence of the truncated unity is benchmarked in Sec.~\ref{sec:tu}.
We conclude in Sec.~\ref{sec:conclusions}.

\section{Model, approximation, and observables}\label{sec:parquet}
We consider the paramagnetic Hubbard model on the square lattice at half-filling,
\begin{align}
    H = &-\sum_{\langle ij\rangle\sigma}{t}_{ij} c^\dagger_{i\sigma}c^{}_{j\sigma}+ U\sum_{i} n_{i\uparrow} n_{i\downarrow},\label{eq:hubbard}
\end{align}
where $t_{ij}$ denotes the hopping between nearest neighbors i and j,
its absolute value $t=1$ sets the unit of energy.
$c^{},c^\dagger$ are the annihilation and creation operators with the spin index $\sigma=\up,\dn$.
We denote the Hubbard repulsion between the densities $n_{\sigma}=c^\dagger_{\sigma}c^{}_{\sigma}$ as $U$;
we consider the weak-coupling regime, $2\leq U/t\leq4$. The lattice size is fixed to $16\times16$.
The temperature is $T/t=0.2$.

We solve the Hubbard model~\eqref{eq:hubbard} using the parquet approximation~\cite{Diatlov57,Dominicis64-2}.
In the following, we recollect only the most essential definitions.
Readers with a background in parquet theory find a complete set of definitions, derivations,
and the calculation cycle of our implementation in Ref.~\cite{Krien21-2}.
The notation used in this work is fully equivalent to Ref.~\cite{Krien21-2},
it corresponds to a compromise between notations frequently used in the parquet and $GW$ literature.
On the other hand, readers more familiar with the fRG find the corresponding definitions in Refs.~\cite{Bonetti22,Walter22},
which use a notation more consistent with the fRG literature.

In the traditional parquet formalism the full vertex function is given in terms of the parquet decomposition,
\begin{align}
F=\Lambda+\Phi^{\ph}+\Phi^{\phv}+\Phi^{\pp}.\label{eq:parquet}
\end{align}
Here, $\Lambda$ is the fully $GG$-irreducible vertex as explained in the introduction.
The $\Phi$'s denote the vertices $GG$-reducible in the horizontal ($\ph$),
vertical ($\phv$), and particle-particle ($\pp$) channel.
Each vertex, e.g., $\Phi^{\ph,\alpha}(k,k',q)$ carries a flavor label,
in the particle-hole channel $\alpha=\ch/\sz$ corresponds charge or spin,
and $k=(\kv,\nu)$, $q=(\qv,\omega)$ denote fermionic, bosonic momentum and Matsubara frequency, respectively.
The parquet decomposition is shown at the bottom of Fig.~\ref{fig:sbe_parquet}.

On the other hand, Refs.~\cite{Krien20,Krien21-2} introduced a bosonized parquet formalism
where vertex diagrams are further decomposed, namely, the full vertex is expressed through the SBE decomposition~\cite{Krien19-2},
\begin{align}
F=\Lambda^\text{Uirr}+\Delta^{\ph}+\Delta^{\phv}+\Delta^{\pp}-2U.\label{eq:sbe_decomposition}
\end{align}
The $\Delta$'s represent the $U$-reducible diagrams which can
be cut in two parts by removing a bare interaction~\cite{Krien21-2,Walter22}.
They are given in terms of the Yukawa coupling (Hedin vertex) and the screened interaction, for example,
\begin{align}
\Delta^{\ph,\alpha}(k,k',q)=\gamma^\alpha(k,q)W^\alpha(q)\gamma^\alpha(k',q).\label{eq:delta}
\end{align}
The bare interaction arises as the leading order of all the $\Delta$'s,
it is therefore subtracted twice in Eq.~\eqref{eq:sbe_decomposition} to avoid overcounting.
Notice that in Eq.~\eqref{eq:sbe_decomposition} it also carries a flavor label, $U^{\ch/\sz}=\pm U$.

In turn, $\Lambda^\text{Uirr}$ is the fully $U$-irreducible vertex given through a parquet-like decomposition,
\begin{align}
\Lambda^\text{Uirr}=\tilde{\Lambda}+M^{\ph}+M^{\phv}+M^{\pp},\label{eq:sbe_parquet}
\end{align}
where $\tilde\Lambda=\Lambda-U$ is the fully $GG$-irreducible vertex with the bare interaction removed.
The $M$'s represent the multi-boson exchange, they are $GG$-reducible but fully $U$-irreducible vertices,
whose momentum-energy dependence does not dissociate in the manner of Eq.~\eqref{eq:delta}.

Inserting Eq.~\eqref{eq:sbe_parquet} into Eq.~\eqref{eq:sbe_decomposition},
the resulting vertex decomposition of the bosonized parquet approach
is shown in Fig.~\ref{fig:sbe_parquet}, above the horizontal line.
It is convenient to add and subtract the bare interaction, represented by a black dot,
so that the diagrams above the horizontal line are arranged consistently:
summing the diagrams in a column yields the corresponding vertex
of the traditional parquet formalism drawn below the horizontal line, for example,
\begin{align}
\Phi^{\ph,\alpha}=\Delta^{\ph,\alpha}-U^\alpha+M^{\ph,\alpha},\;\alpha=\ch,\sz,\label{eq:phi}
\end{align}
which connects the traditional and the bosonized parquet quantities on the left- and right-hand-side, respectively.
Lastly, we introduce the parquet approximation:
\begin{align}
\tilde{\Lambda}\equiv0.\label{eq:parquet_approximation}
\end{align}
As a matter of fact, this is a rich approximation with nontrivial properties, as our results exemplify.

In this work we consider only the particle-hole quantities in Eqs.~\eqref{eq:delta} and~\eqref{eq:phi}.
In the traditional parquet formalism the set of equations is closed via the Bethe-Salpeter,
Dyson, and Schwinger-Dyson equations.
In the bosonized formalism the latter is replaced with the Hedin equation ($\Sigma=GW\gamma$) 
and one defines a bosonic self-energy ($\Pi=GG\gamma$),
which determines the screened interaction via another Dyson-like equation ($W=U+U\Pi W$).
However, to keep the presentation concise and general, we refer to Refs.~\cite{Krien21-2,Bonetti22,Walter22}
for detailed information including a calculation cycle or (m)fRG flow equations, respectively.

In our numerical application we use $N^\gamma_\nu=32$ fermionic and $N^\gamma_\omega=32$
bosonic Matsubara frequencies for the Yukawa couplings $\gamma$.
The $M$'s are evaluated on a smaller fermionic frequency grid, using $N^M_\nu=16$ and  $N^M_\omega=32$ for bosonic frequencies.
Even though frequency summations like $\sum_{\nu'} M(\nu,\nu',\omega)G(\nu')G(\nu'+\omega)$ decay by one power of $\nu'$
faster compared to a summation over the corresponding $\Phi$'s of the traditional parquet approach,
a cutoff error arises in the $\gamma$'s for $\nu\approx (2N^M_\nu+1)\pi T$.
Using smaller momentum grids we checked that our results for small $\nu$ presented in the following
are not affected qualitatively by the frequency cutoff error. Quantitative convergence analysis for the $16\times16$ grid is however beyond computational capability of the current implementation.
As $\gamma$ determines the key observables $G$ and $W$, it is desirable to achieve convergence in $\gamma$ with respect to frequencies which would
correspond to a very high standard of convergence for the parquet approach.
An asymptotic treatment of $\gamma$ goes however beyond the established theory of vertex asymptotics~\cite{Wentzell20};
this problem may be considered elsewhere in the future.

\begin{figure*}
  \begin{center}
    \begin{tikzpicture}
    \node[anchor=south west,inner sep=0] (imagef0) at (0,0) {\includegraphics[width=\textwidth]{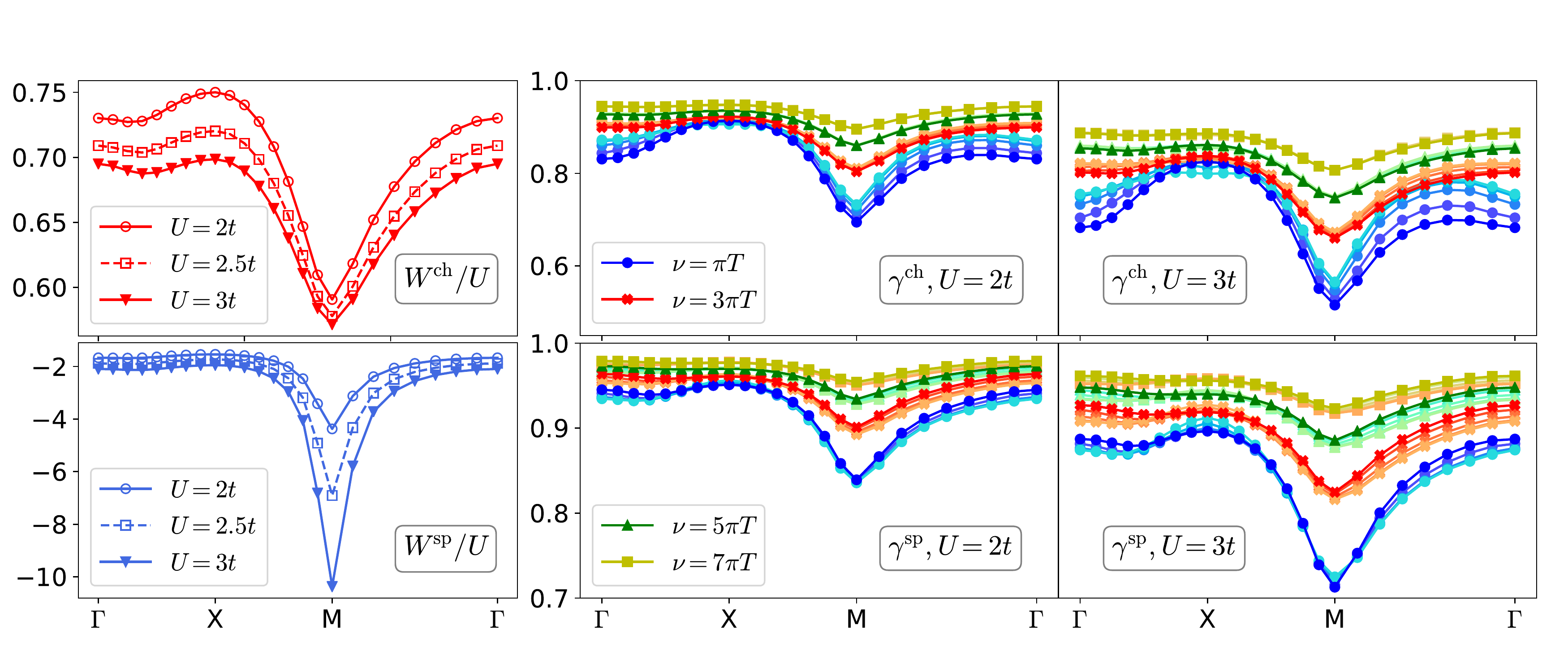}};
    \end{tikzpicture}
\end{center}
    \caption{\label{fig:wq} Left: The screened interaction in the charge (red) and spin (blue) channel normalized by the bare interaction. Center (right): Yukawa coupling in the charge
    (spin) channel as a function of $\qv$. Different palettes
    show the first four fermionic frequencies $\nu$, color
    tones indicate momenta $\kv$ (see text).
    }
\end{figure*}

\section{Screened interaction and Yukawa coupling}\label{sec:wq}
Fermionic properties of the Hubbard model at weak coupling,
in particular the formation of a pseudogap due to long-ranged spin fluctuations,
have been discussed in great detail in the recent literature, see, for example, Refs.~\cite{Hille20-2,Schaefer21}.
However, electronic correlations renormalize also the Yukawa coupling between fermions and bosons,
an effect which has received much less attention~\cite{Krien20-2}.
The parquet approach respects the crossing symmetry and hence provides us by construction with the full dependence
of the Yukawa couplings on fermionic and bosonic momentum.
Notice that we do not enter the pseudogap regime,
which requires roughly 1000 lattice sites to avoid a finite-size effect~\cite{Schaefer21}.
However, we still observe an interesting evolution of $\gamma$ as antiferromagnetic fluctuations begin to build up.

\subsection{Screened interaction}\label{sec:w}
The left panels of Fig.~\ref{fig:wq} show the static screened interactions $W^{\ch/\sz}(\qv,\omega=0)$
along the high-symmetry path. For comparison we normalize it with the absolute value $U$ of the bare interaction.
The sign of the different curves therefore signals repulsion ($W/U>0$) or attraction ($W/U<0$)
and the amplitude indicates whether the interaction $U^{\ch/\sz}=\pm U$ in the respective
channel is screened ($|W/U|<1$) or enhanced ($|W/U|>1$).
As expected, with increasing $U$ a strong attractive interaction develops
in the spin channel along the $\qv=(\pi,\pi)$ direction.

\subsection{Yukawa couplings}\label{sec:gamma}
The center and right panels show the Yukawa coupling $\gamma^{\ch/\sz}(\kv,\nu,\qv,\omega=0)$
between fermions and static charge/spin fluctuations as a function of the bosonic momentum $\qv$ for $U/t=2$ and $U/t=3$.
The four color palettes (blue, red, green, yellow) correspond to the four smallest fermionic Matsubara frequencies
($\nu_0=\pi T, \nu_1=3\pi T, \nu_2=5\pi T, \nu_3=7\pi T$), respectively.
Colors from dark to light indicate fermionic momenta $\kv=(x,\pi-x)$ on the Fermi surface,
where $x\in\{0,\pi/8,\pi/4,3\pi/8,\pi/2\}$,
starting with the antinode $(0,\pi)$ [darkest] and ending with the node $(\pi/2,\pi/2)$ [lightest].
Notice that at particle-hole symmetry the $\gamma$'s are real-valued.

Overall, $\gamma^{\ch/\sz}$ depend most strongly on $\qv$, less strongly on $\nu$,
and the least on $\kv$ (dependence on $\omega$ will be considered elsewhere).
However, this can not be generalized, as $\gamma^{\ch}(\kv,\nu=\pi T,\qv,\omega=0)$
shows a sizable $\kv$-dependence for $U/t=3$, 
whereas $\gamma^{\sz}$ is largely independent of $\kv$ for the same set of parameters and labels.
In the non-interacting system the Yukawa coupling is unity;
Fig.~\ref{fig:wq} shows that a weak interaction leads to screening ($\gamma^{\ch/\sz}<1$).
Notice that $\gamma$ determines both the fermionic ($\Sigma=GW\gamma$),
as well as the bosonic self-energy ($\Pi=GG\gamma$), which also enters $\Sigma$ via $W$.
Close to an instability an increase of $\gamma$, even by a few percent, can drastically enhance $W$.
Indeed, we showed recently that even for the harmless parameters $U/t=2, T/t=0.2$
the screening of $\gamma^\sz$ is indispensable to obtain a reasonable approximation for $\Sigma$~\cite{Krien21-2}.

Furthermore, as the system is driven to the antiferromagnetic instability, fermions decouple from the soft bosons
($\gamma^\sz\rightarrow0$ as $W^\sz\rightarrow-\infty$),
since the Goldstone excitations of the ordered phase are protected (Adler principle,~\cite{Adler65,Schrieffer95}).
Indeed, comparing $U/t=2$ and $U/t=3$ in Fig.~\ref{fig:wq}
we see that $\gamma^\sz$ is much more strongly screened around $\qv=(\pi,\pi)$ for the larger interaction,
which corresponds to a longer correlation length {(see also Sec.~\ref{sec:tu})}.
On the other hand, we found in recent investigations that, as soon as fermionic states are destroyed due
to the feedback from the spin fluctuations, this requirement is lifted and $\gamma^\sz$ rises again for those
$\kv$ where a pseudogap opens, resulting in a nodal/antinodal dichotomy
of $\gamma^\sz$ with respect to $\kv$~\cite{Krien20-2,Krien21}.
There hence exists a subtle interplay between bosonic fluctuations, Fermi surface features, and the Yukawa couplings,
which needs to be considered when dependencies of the latter are neglected or parametrized.

\begin{figure}
  \begin{center}
    \begin{tikzpicture}
    \node[anchor=south west,inner sep=0] (imagef0) at (0,0) {\includegraphics[width=0.5\textwidth]{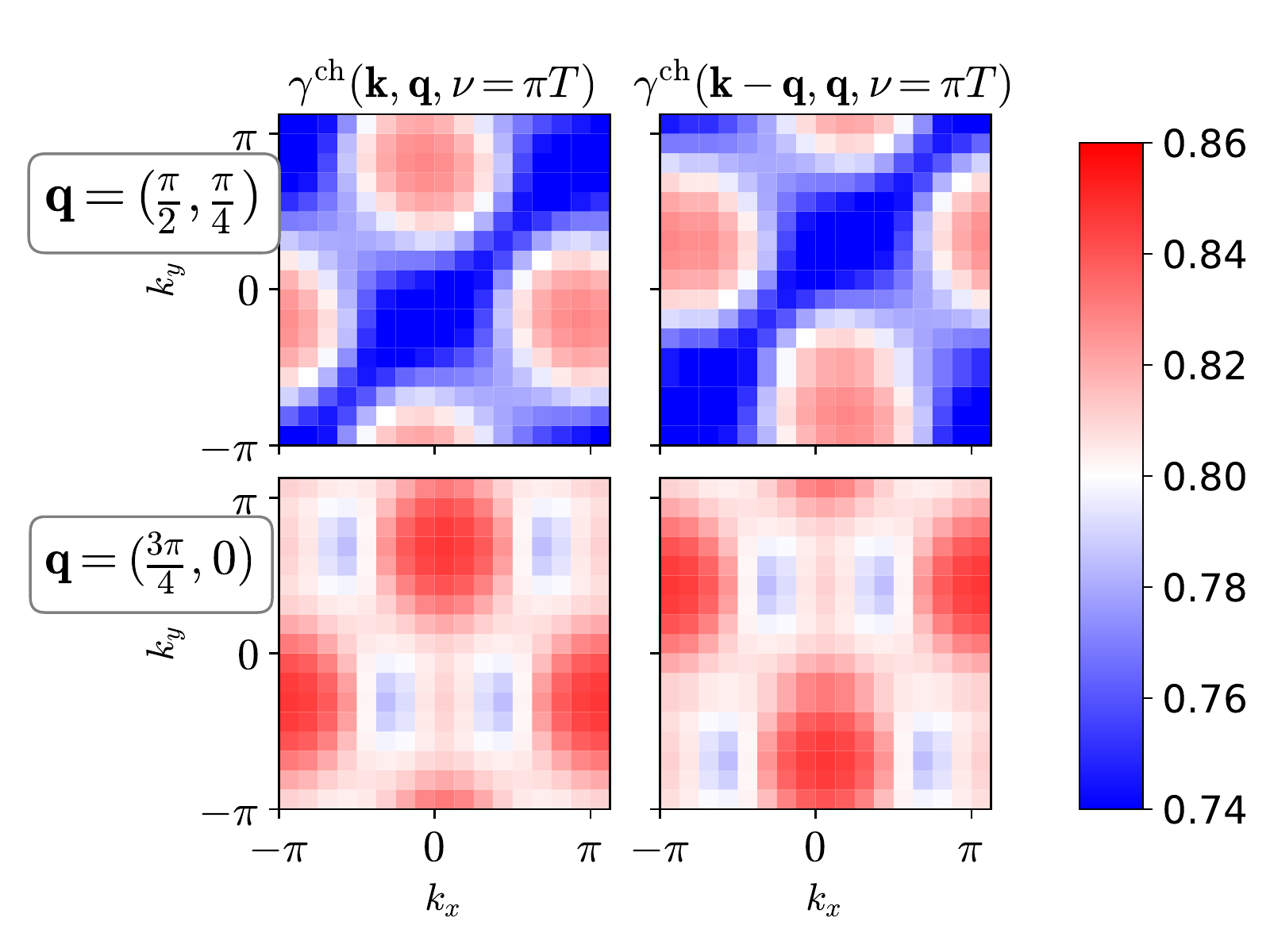}};
    \end{tikzpicture}
\end{center}
    \caption{\label{fig:sym} Numerical validation of Eq.~\eqref{eq:sym} for the static charge Yukawa coupling
    $\gamma^\ch(\kv,\qv,\nu,\omega=0)$: shifting $\kv$ by $-\qv$ is the same as going from $\kv$ to $-\kv$.
    }
\end{figure}

\subsection{Symmetries}
We also discuss symmetries of the Yukawa couplings, see Refs.~\cite{vanLoon18,Rohringerthesis}.
Firstly, we note that inversion symmetry of the lattice,
as well as time-reversal and SU($2$) symmetry are required
for the derivation in Ref.~\cite{Krien21-2} and by our implementation.
This set of symmetries allows to interchange the fermionic labels of the full vertex function $F(k,k',q)=F(k',k,q)$,
see also Refs.~\cite{Rohringer12,Rohringerthesis}. Since the Yukawa coupling is just a four-point vertex with tapered
Green's function legs on one side (plus $1$)~\cite{Krien21-2}, the symmetry of the full vertex implies that it does not matter on which side the legs are attached.
As a result, the left- and right-handed Yukawa couplings shown in Fig.~\ref{fig:sbe_parquet} are identical.
It is important to keep in mind, however, that in a more general setting our formalism needs to be re-derived using left- and right-handed Yukawa couplings~\cite{Bonetti22,Walter22}.

A symmetry valid by definition is due to complex conjugation, $\gamma^*(k,q)=\gamma(-k,-q)$.
On the other hand, the $\gamma$'s are invariant under symmetry operations
of the point group of the lattice~\cite{Thomale13}. For example, inversion symmetry implies
$\gamma(\kv,\qv,\nu,\omega)=\gamma(-\kv,-\qv,\nu,\omega)$.
Since the symmetry operations needs to be applied to both momenta at the same time,
in a practical implementation only one of the momenta can be mapped to the irreducible wedge of the lattice.
Hence, for the $16\times16$ square lattice each Yukawa coupling requires $\sim\frac{45}{256}(256)^2 N^\gamma_\nu N^\gamma_\omega$ complex numbers. Inversion combined with complex conjugation further implies $\gamma(\kv,\qv,\nu,\omega)=\gamma^*(\kv,\qv,-\nu,-\omega)$.
Since $\gamma$ is real-valued at particle-hole symmetry it follows for $\omega=0$ that
\begin{align}
\gamma(\kv,\qv,\nu,\omega=0)=&\gamma(\kv,\qv,-\nu,\omega=0),\label{eq:gamma_minus}
\end{align}
which we use in the following section.

\begin{figure}[b]
    \begin{center}
	\begin{tikzpicture}%[scale=1]
        \begin{scope}[scale=1.3]
        %\draw[thick,middlearrow={>}] (-1,0) -- (0,-1) -- (1,0) -- (0,1) -- cycle;
        \draw[thick,middlearrow={>}] (-1,0) -- (0,-1);
        \draw[thick,middlearrow={>}] (0,-1) -- (1,0);
        \draw[thick,middlearrow={>}] (1,0) -- (0,1);
        \draw[thick,middlearrow={>}] (0,1) -- (-1,0);
        \node at (-1.6,0) {$(-\pi,0)$};
        \node at (1.5,0) {$(\pi,0)$};
        \node at (0,-1.3) {$(0,-\pi)$};
        \node at (0,1.3) {$(0,\pi)$};
        \node at (-.7,0) {$1$};
        \node at (.7,0) {$3$};
        \node at (0,-.7) {$2$};
        \node at (0,.7) {$4$};
        \draw[thick] (1,1) -- (1,-1) -- (-1,-1) -- (-1,1) -- cycle;
        %\draw[->] (-1+.2,0) -- (.2,-1+.2);
        \end{scope}
    \end{tikzpicture}
\end{center}
    \caption{\label{fig:path} Path on the Fermi surface traversed by $\kv_F, \kv'_F$.
    }
\end{figure}
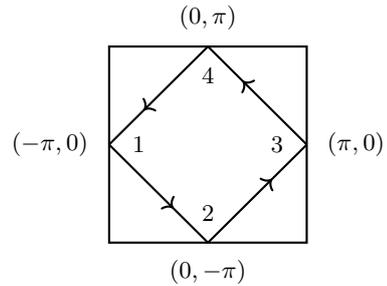

Lastly, we verify numerically that a nontrivial symmetry of the $\gamma$'s is respected by our implementation.
Namely, the full four-point vertex satisfies by definition the ``swapping symmetry'' $F_{kk'q}=F_{k'+q,k+q,-q}$~\cite{Galler17}.
Together with $F_{kk'q}=F_{k'kq}$ it follows~\cite{vanLoon14} that
${\gamma}^{\ch/\sz}(k-q,q)={\gamma}^{\ch/\sz}(k,-q)$~\footnote{
For completeness, we report also the corresponding symmetry for the singlet particle-particle vertex~\cite{Krien21-2}:
${\gamma}^{\sing}_{k+q,q}={\gamma}^{\sing}_{-k,q}$.
We do not consider this vertex here since at particle-hole symmetry it can be obtained from the charge vertex, $\gamma^\sing(k,q)=-\gamma^\ch(-k,q)$~\cite{Krien19-2}.}.
We set $q=(\qv,\omega=0)$, resulting in,
\begin{align}
&{\gamma}^{\ch/\sz}(\kv-\qv,\qv,\nu,\omega=0)\nonumber\\
=&{\gamma}^{\ch/\sz}(\kv,-\qv,\nu,\omega=0)\nonumber\\
=&{\gamma}^{\ch/\sz}(-\kv,\qv,\nu,\omega=0).\label{eq:sym}
\end{align}
In the last line we applied the inversion symmetry.
Equation~\eqref{eq:sym} implies for ${\gamma}^{\ch/\sz}(\kv,\qv,\nu,\omega=0)$ that shifting $\kv\rightarrow\kv-\qv$
has the same effect as $\kv\rightarrow-\kv$. That this is indeed the case in our implementation can be seen
in Fig.~\ref{fig:sym} which shows $\gamma^\ch$ for, e.g., $\qv=(\pi/2,\pi/4)$ and $\qv=(3\pi/4,0)$.
We chose here $\gamma^\ch$ for $U/t=3$, as it depends strongly on $\kv$ (see Fig.~\ref{fig:wq}),
and incommensurate $\qv$ for a generic result. Symmetries put strong conditions on the $\gamma$'s
which are useful to verify code during debugging, or to save memory space.

\section{Single- and multi-boson exchange}\label{sec:sbembe}
We analyze the quantities $\Phi, \Delta,$ and $M$ in Eq.~\eqref{eq:phi}.
These are four-point vertex functions depending on three momenta $\kv,\kv',\qv$,
and three frequencies $\nu,\nu',\omega$.
To get a grasp of these quantities, we focus on fermionic momenta $\kv_F, \kv'_F$ on the Fermi surface
which traverse the path shown in Fig.~\ref{fig:path}, thereby passing through all four antinodal points.
The fermionic frequencies are set to $\nu=\nu'=\pi T$ or $\nu=-\nu'=-\pi T$.
We focus on the static limit $\omega=0$ and first set the bosonic transfer momentum to $\qv=(\pi,\pi)$,
which always guides scattered quasiparticles to final states on the Fermi surface.

\begin{figure}
  \begin{center}
    \begin{tikzpicture}
    \node[anchor=south west,inner sep=0] (imagef0) at (0,0) {\includegraphics[width=0.5\textwidth]{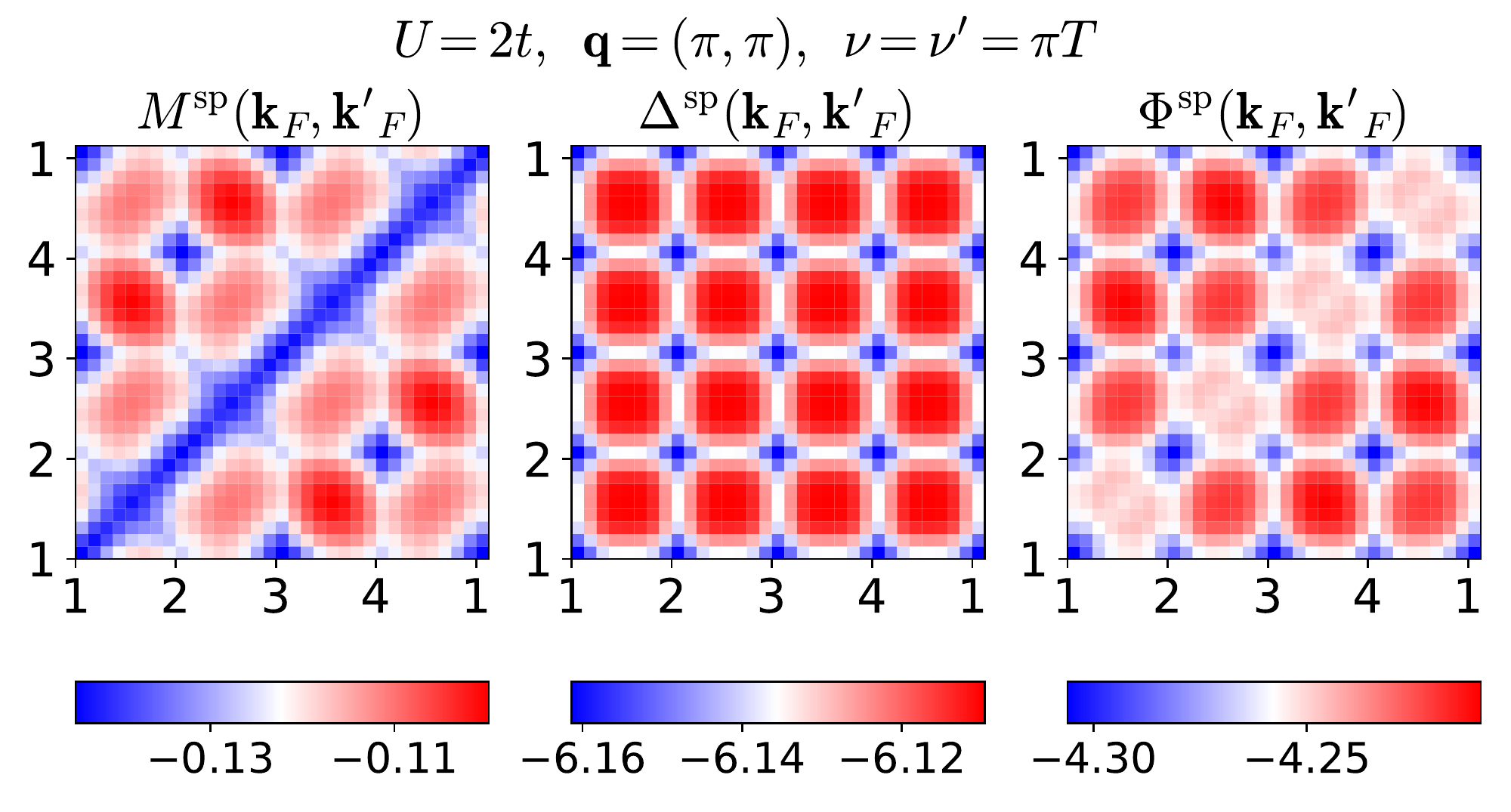}};
    \node[anchor=south west,inner sep=0] (imagef0) at (0,-5.0) {\includegraphics[width=0.5\textwidth]{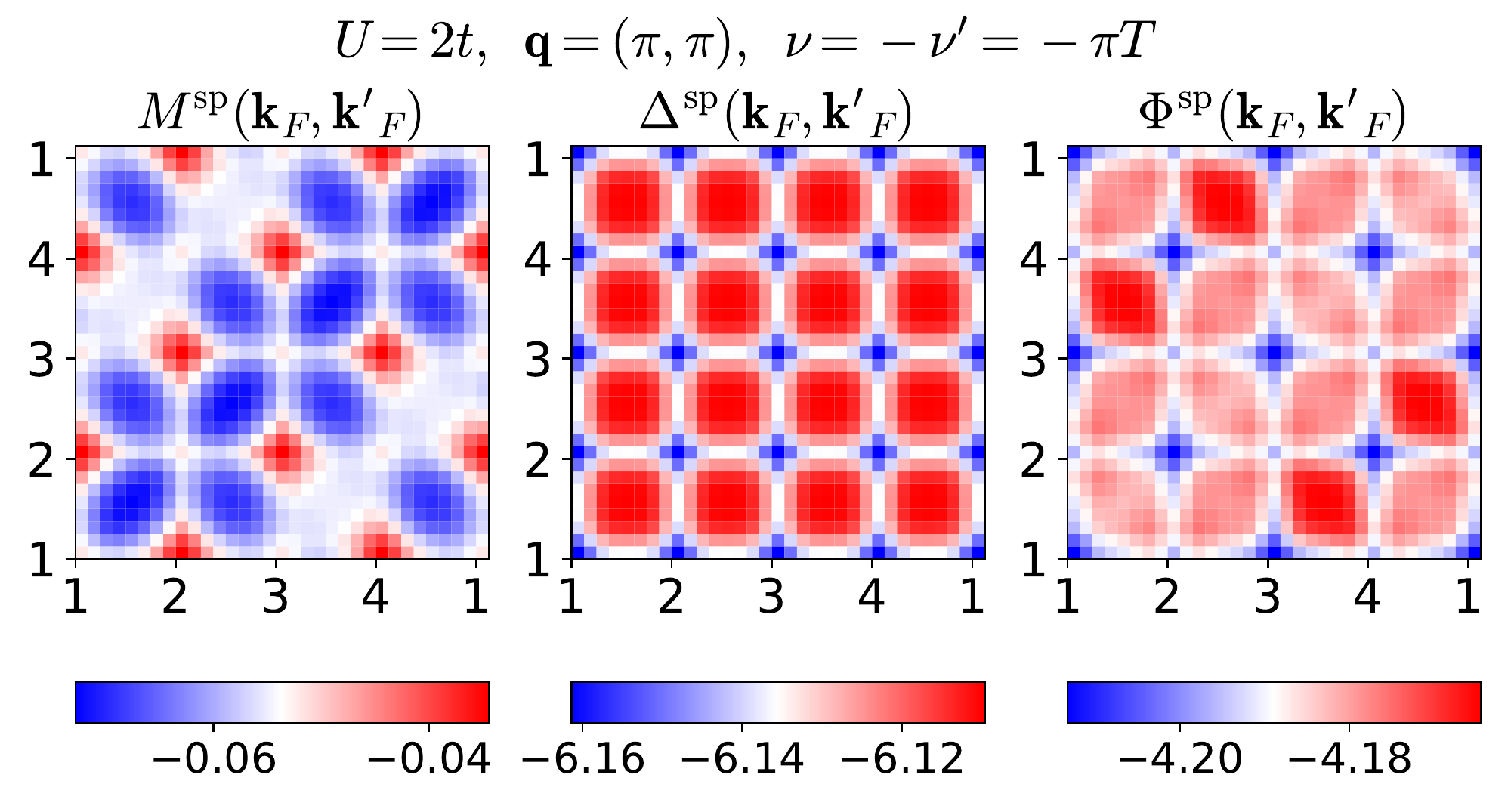}};
    \end{tikzpicture}
\end{center}
    \caption{\label{fig:u2spqpi} Momentum dependence of spin multi- (left) and single-boson (center)
    exchange for $U/t=2$. Right panels show the corresponding
    reducible vertex of the traditional parquet formalism.
    Top (bottom) panels show $\nu=\pi T$ ($\nu=-\pi T$).
    The two fermionic momenta traverse along the entire Fermi surface as shown in Fig.~\ref{fig:path},
    other labels as shown in the title.}
\end{figure}

\begin{figure}
  \begin{center}
    \begin{tikzpicture}
    \node[anchor=south west,inner sep=0] (imagef0) at (0,0) {\includegraphics[width=0.5\textwidth]{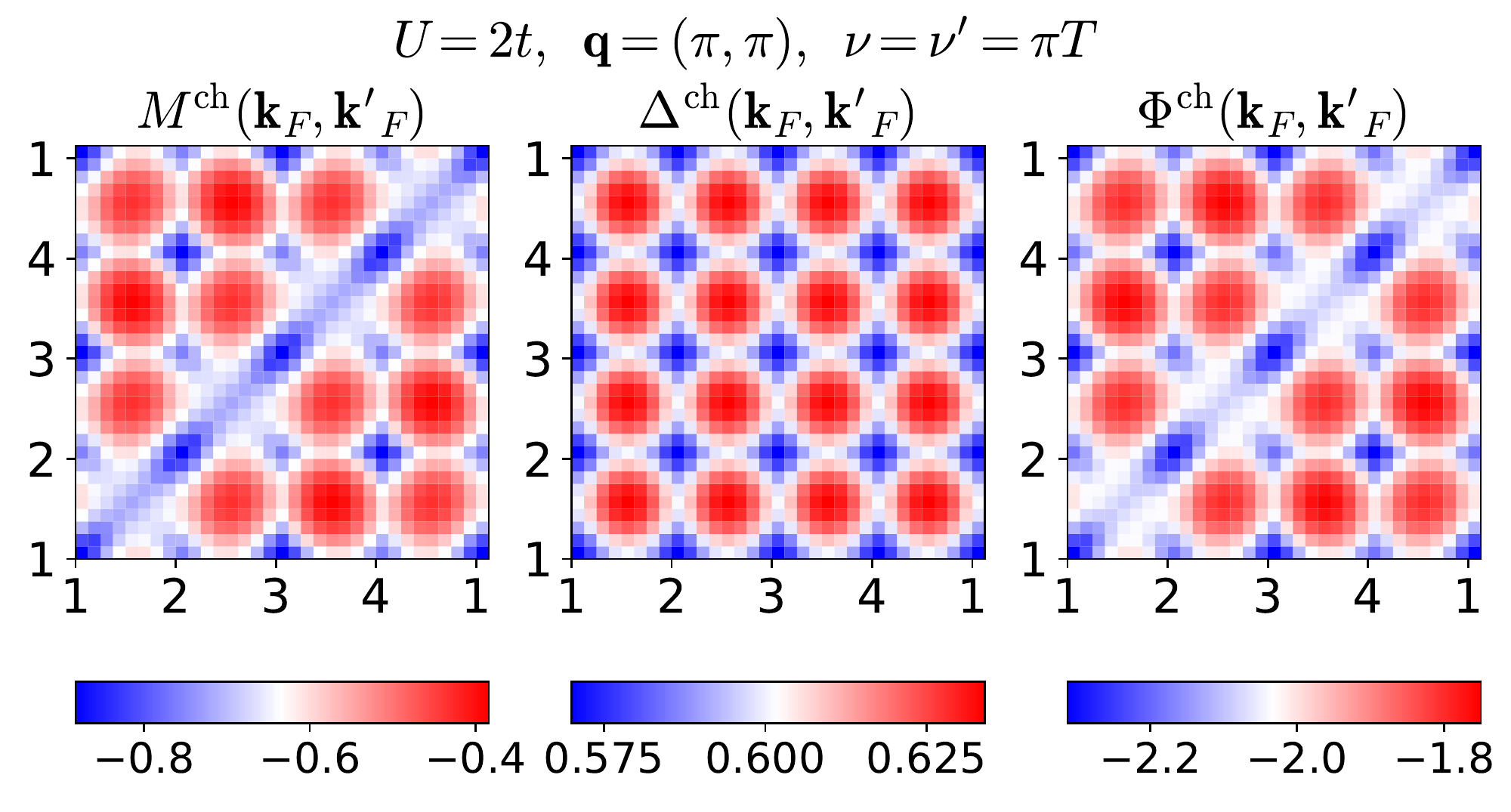}};
    \node[anchor=south west,inner sep=0] (imagef0) at (0,-5.0) {\includegraphics[width=0.5\textwidth]{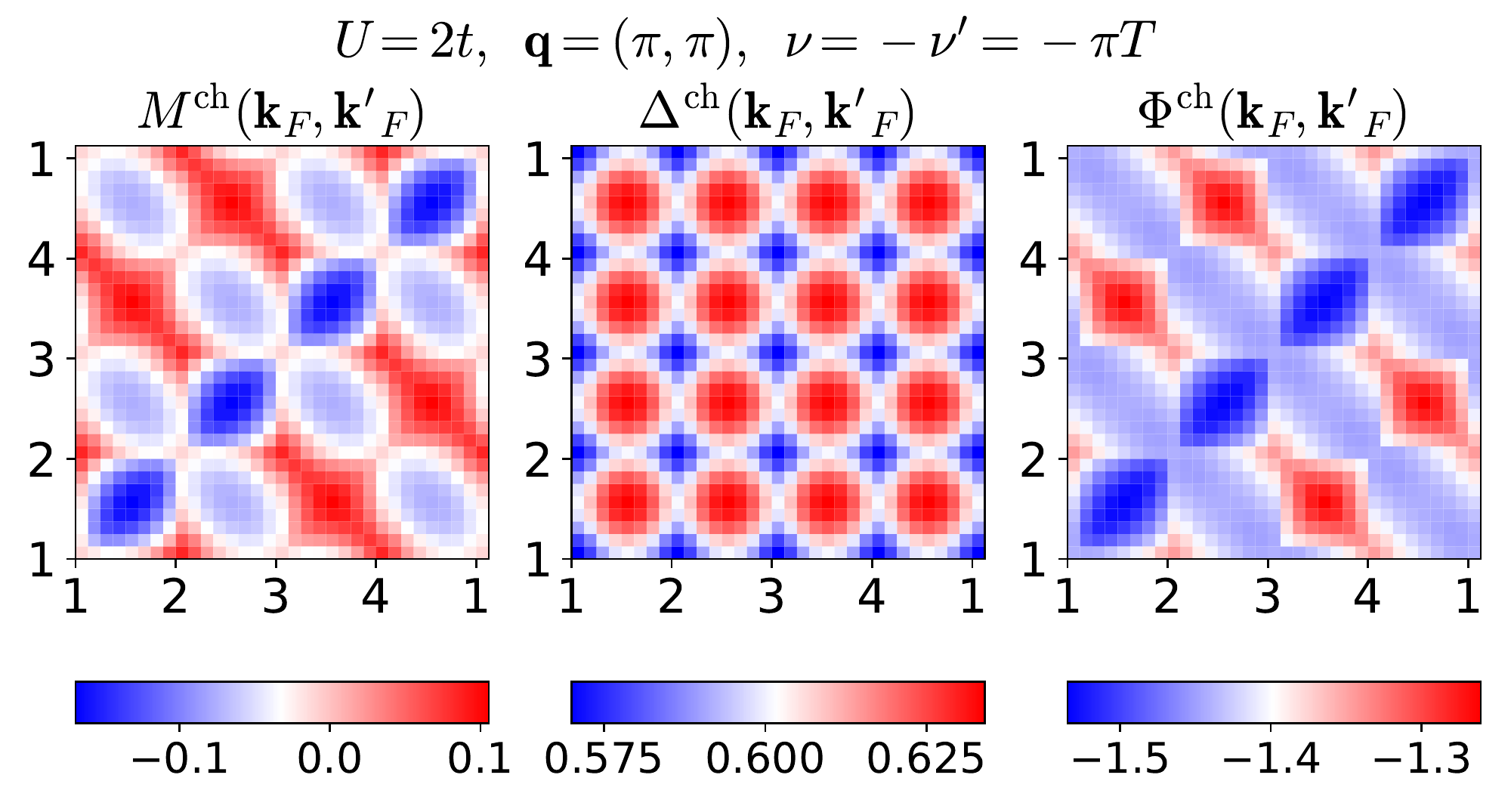}};
    \end{tikzpicture}
\end{center}
    \caption{\label{fig:u2chqpi} Charge quantities corresponding to Fig.~\ref{fig:u2spqpi}.}
\end{figure}

In this manner we plot $M^\sz(\kv_F,\kv_F',\qv=(\pi,\pi),\nu=\pi T,\nu'=\pi T,\omega=0)$
for $U/t=2$ in the top left panel of Fig.~\ref{fig:u2spqpi}. 
Comparison with $\Delta^\sz$ with the same labels, drawn in the center,
shows that the latter exhibits a higher symmetry with respect to the fermionic momenta.
Finally, $\Phi^\sz$ on the right is obtained as the sum of $M^\sz$ and $\Delta^\sz$,
with the bare interaction $U^\sz=-U$ subtracted
[cf. Eq.~\eqref{eq:phi} and compare the magnitude of the color bars]. The high symmetry of $\Delta^\sz$,
which repeats along each of the four edges of the Fermi surface (cf. Fig.~\ref{fig:path}),
implies that in a scattering event of two quasiparticles,
mediated by this vertex, it is irrelevant to which of the four edges their initial momenta belong.
In contrast, the lower symmetry of $M^\sz$ implies that it mediates scattering events
where it {\sl does} matter whether the respective scattering partner
lives on the same, an adjacent, or on the opposite edge of the Fermi surface.

Let us now consider the effect of flipping the sign of one fermionic frequency, $\nu\rightarrow-\pi T$.
According to Eq.~\eqref{eq:gamma_minus} in the previous section, $\gamma^\sz(\omega=0)$ is symmetric
with respect to $\nu$. Since the frequency dependence of the
$\Delta$'s stems from the $\gamma$'s, $\Delta(\omega=0)$ is also invariant under the sign flip,
which can be seen in the bottom center panel of Fig.~\ref{fig:u2spqpi}.
The situation is again quite different for $M^\sz$ whose momentum structure is completely overturned
under the sign flip of $\nu$. It was observed already in Refs.~\cite{Krien19-2,Harkov21} that
the fully $U$-irreducible vertex changes drastically when going from the sectors
$\text{sgn}(\nu)=\text{sgn}(\nu')$ to $\text{sgn}(\nu)=-\text{sgn}(\nu')$.
Apparently, in case of nonlocal correlations this is intertwined with its dependence on the fermionic momenta.

The patterns visible in $\Phi^\sz$ arise from the superposition of those in $M^\sz$ with
the more symmetric ones in $\Delta^\sz$, with an optically astounding result.
Notice however that the color plot overemphasizes small variations in these quantities.
It is $|\Delta^\sz|\gg|M^\sz|$, because the former inherits a large absolute value from
$W^\sz(\qv=(\pi,\pi),\omega=0)$, and a weak $\kv$ dependence from $\gamma^\sz$ (cf. Fig.~\ref{fig:wq}).
We find that for larger interaction the difference in magnitude is even more enhanced
and a discussion of the tiny variations is moot.

\begin{figure}
  \begin{center}
    \begin{tikzpicture}
    \node[anchor=south west,inner sep=0] (imagef0) at (0,0) {\includegraphics[width=0.5\textwidth]{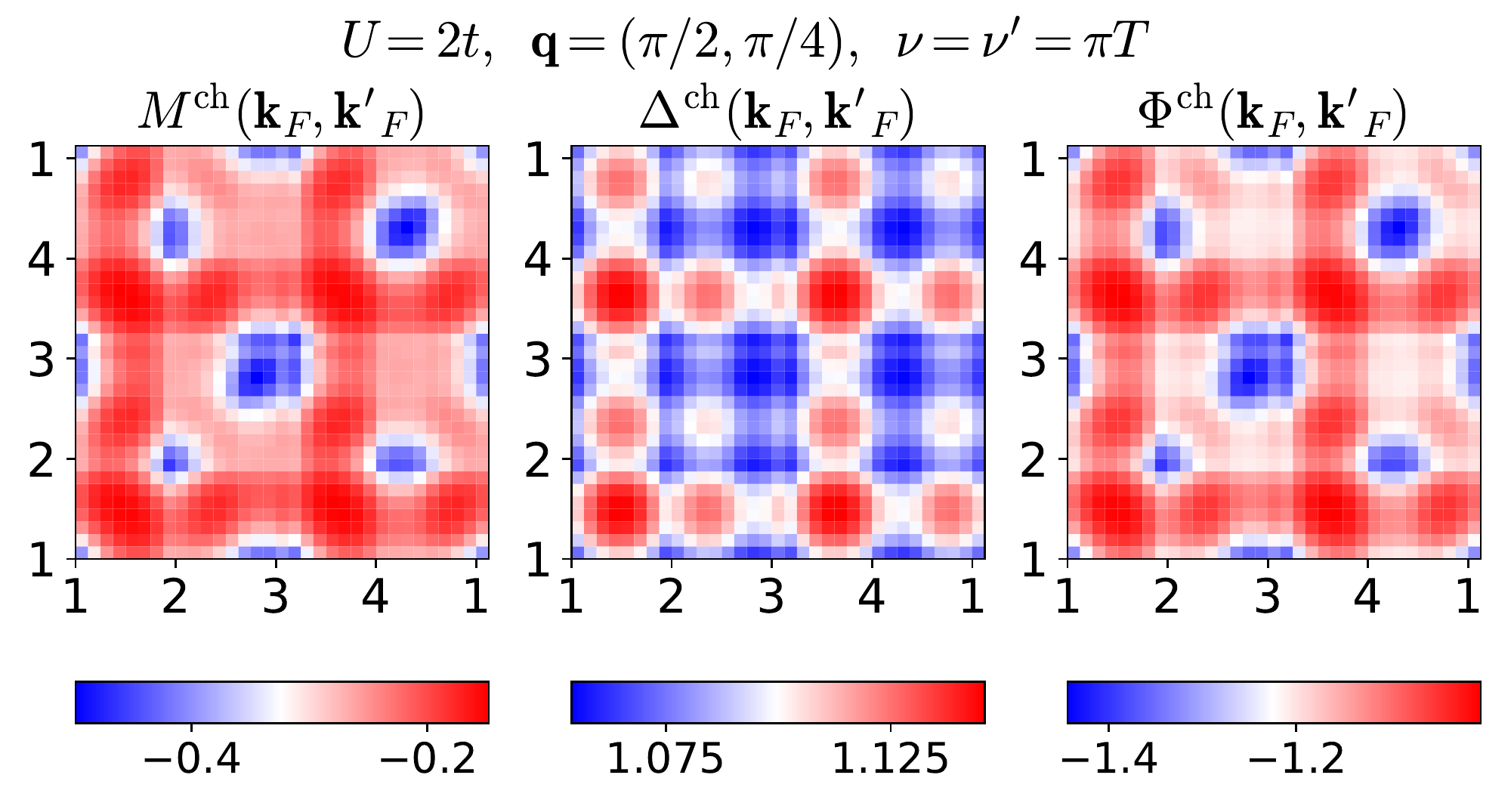}};
    \node[anchor=south west,inner sep=0] (imagef0) at (0,-5.0) {\includegraphics[width=0.5\textwidth]{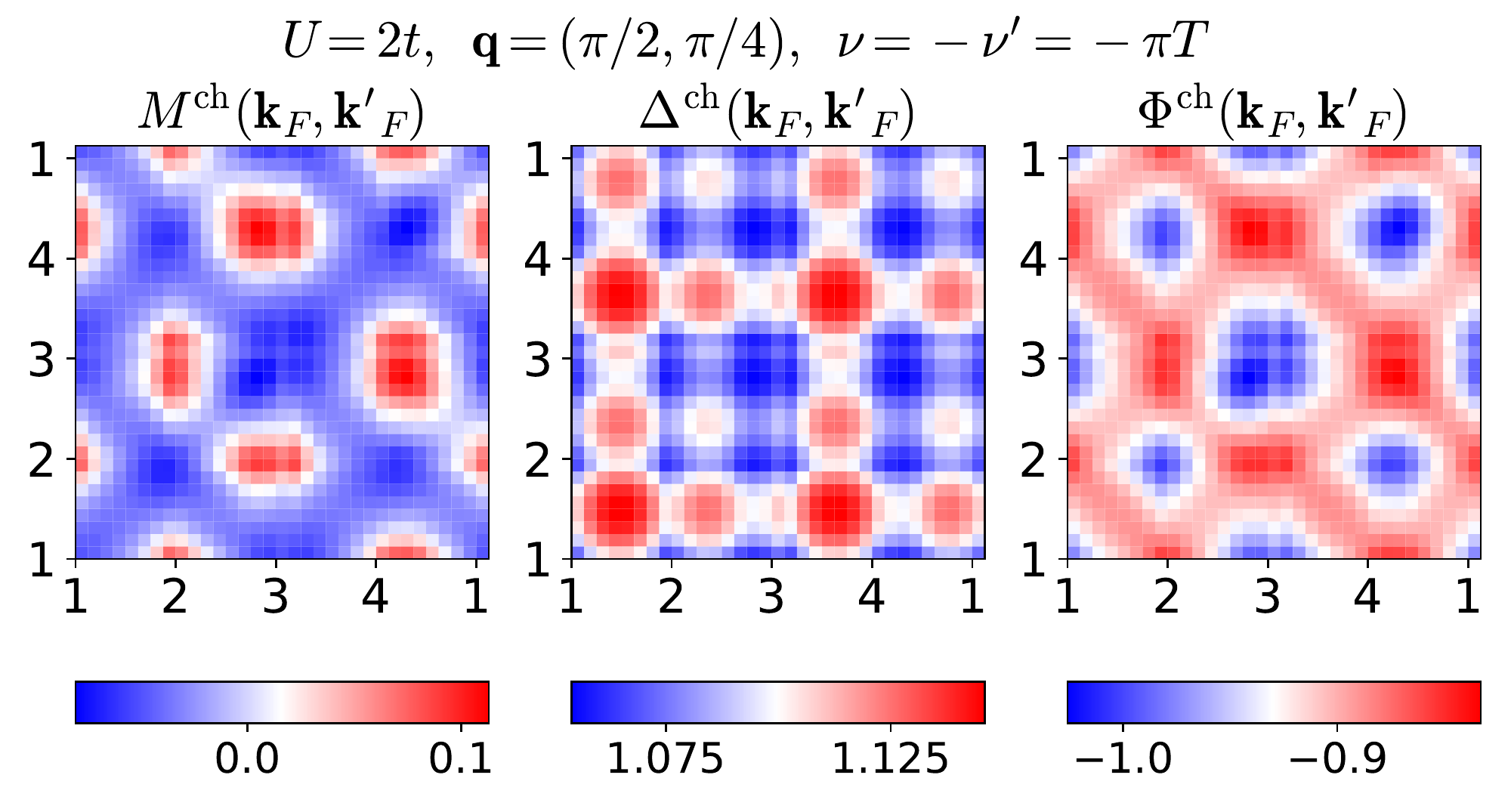}};
    \end{tikzpicture}
\end{center}
    \caption{\label{fig:u2qpi2pi4} Charge quantities as in Fig.~\ref{fig:u2chqpi} for incommensurate
    bosonic momentum $\qv=(\pi/2,\pi/4)$.}
\end{figure}

However, in the charge channel we find that $M^\ch$ is larger than $\Delta^\ch$ at small frequencies, see Fig.~\ref{fig:u2chqpi}. 
The resulting $\Phi^\ch$ is thus dominated by $M^\ch$.
Again $\Delta^\ch$ is symmetric with respect to momenta
and under a sign flip of $\nu$, whereas $M^\ch$ not only changes its asymmetric momentum structure
completely under the sign flip, but also its magnitude by a factor $4$ to $8$.
Finally, we also present the charge quantities for an incommensurate bosonic momentum,
$\qv=(\pi/2,\pi/4)$, in Fig.~\ref{fig:u2qpi2pi4}.
Although $\Delta^\ch$ retains some regularity compared to $M^\ch$,
it loses much of its symmetry with respect to momenta, but remains symmetric under under a sign flip of $\nu$.

\section{Truncated unity and vertex asymptotics}\label{sec:tu}
\subsection{Convergence of the truncated unity}
While in this work we kept the full momentum dependence of the various vertex functions,
this is in general undesirable beyond applications to simple model systems.
It is therefore, on one hand, a question of practical interest to parametrize the momentum dependencies
in a memory-efficient way. On the other hand, the formal construction of the theory should also work towards this goal. Here, for example, the single-boson exchange $\Delta$
is by construction parametrized through $W$ and $\gamma$.
However, if a simplified parquet or fRG scheme keeps also the multi-boson exchange $M$,
the question arises whether the bosonized theory offers any advantages over a
traditional fermionic formulation using the $\Phi$'s.
Moreover, the vertex asymptotics~\cite{Wentzell20} is often used to parametrize the $\Phi$'s at high frequencies.
Since the vertex asymptote corresponds itself to high-frequency limits of the $\Delta$'s~\cite{Krien19-2,Harkov21},
the bosonized theory may only offer advantages in the low-frequency regime.

In this regard, Ref.~\cite{Harkov21} recently demonstrated that the $\Delta$'s capture resonant
low-frequency features of the local full vertex function $F_\text{loc}$ of the Anderson impurity model (AIM).
Even though other low-frequency features reside in the $M$'s,
the two-particle quantities of the DMFT approximation are recovered to good accuracy using only the $\Delta$'s
(cf. Fig.~\ref{fig:sbe_parquet}; $F_\text{loc}$ was approximated by neglecting all of the $M$'s).
If however low-frequency information in the $\Delta$'s is also neglected,
the parametrization of $F_\text{loc}$ fails at strong coupling~\cite{Harkov21}.
Concretely, we find the difference between $\Delta^\sz$ and its asymptotic expression as follows,
\begin{align}
\Delta^\sz(k,k',q)=&W^\sz(q)\left[\gamma^\sz(k,q)+\gamma^\sz(k',q)-1\right]\nonumber\\
&+\Delta_R^\sz(k,k',q)\nonumber\\
\Delta_R^\sz(k,k',q)=&[\gamma^\sz(k,q)-1]W^\sz(q)[\gamma^\sz(k',q)-1]\label{eq:deltar},
\end{align}
and Ref.~\cite{Harkov21} showed for the AIM ($k,k',q\rightarrow\nu,\nu',\omega$)
that an approximation for $F_\text{loc}$ should keep the term $\Delta^\sz_R$,
which vanishes asymptotically for $|\nu|\rightarrow\infty$ and/or $|\nu'|\rightarrow\infty$.

Here we draw an analogy to the present investigation:
While the effective AIM of the DMFT approximation exhibits strong local spin fluctuations at strong coupling,
here the Hubbard model at weak coupling develops long-ranged spin-density wave fluctuations.
Physically these scenarios are of course quite different;
for example, in the AIM $\gamma^\sz$ seems to diverge for small $\nu$ and low temperature~\cite{Harkov21},
while Fig.~\ref{fig:wq} shows that in the Hubbard model $0<\gamma^\sz<1$ is screened.
However, a similarity is that the screened interaction $W^\sz$ is large, either due to the local moment in the AIM, or, here, {because of the growing antiferromagnetic correlation length $\xi$.}
In the latter case it is therefore plausible that the term $\Delta_R^\sz$ in
Eq.~\eqref{eq:deltar} grows with $\xi$,
and at the same time also develops a strong dependence on the bosonic momentum $\qv$.
In this case it could be advantageous to keep $\Delta_R^\sz$ parametrized as a part of $\Delta^\sz$,
rather than to assign it to a memory-intensive four-point vertex.
This is what we show in the following.

\begin{figure}
  \begin{center}
    \begin{tikzpicture}
    \node[anchor=south west,inner sep=0] (imagef0) at (0,0) {\includegraphics[width=0.5\textwidth]{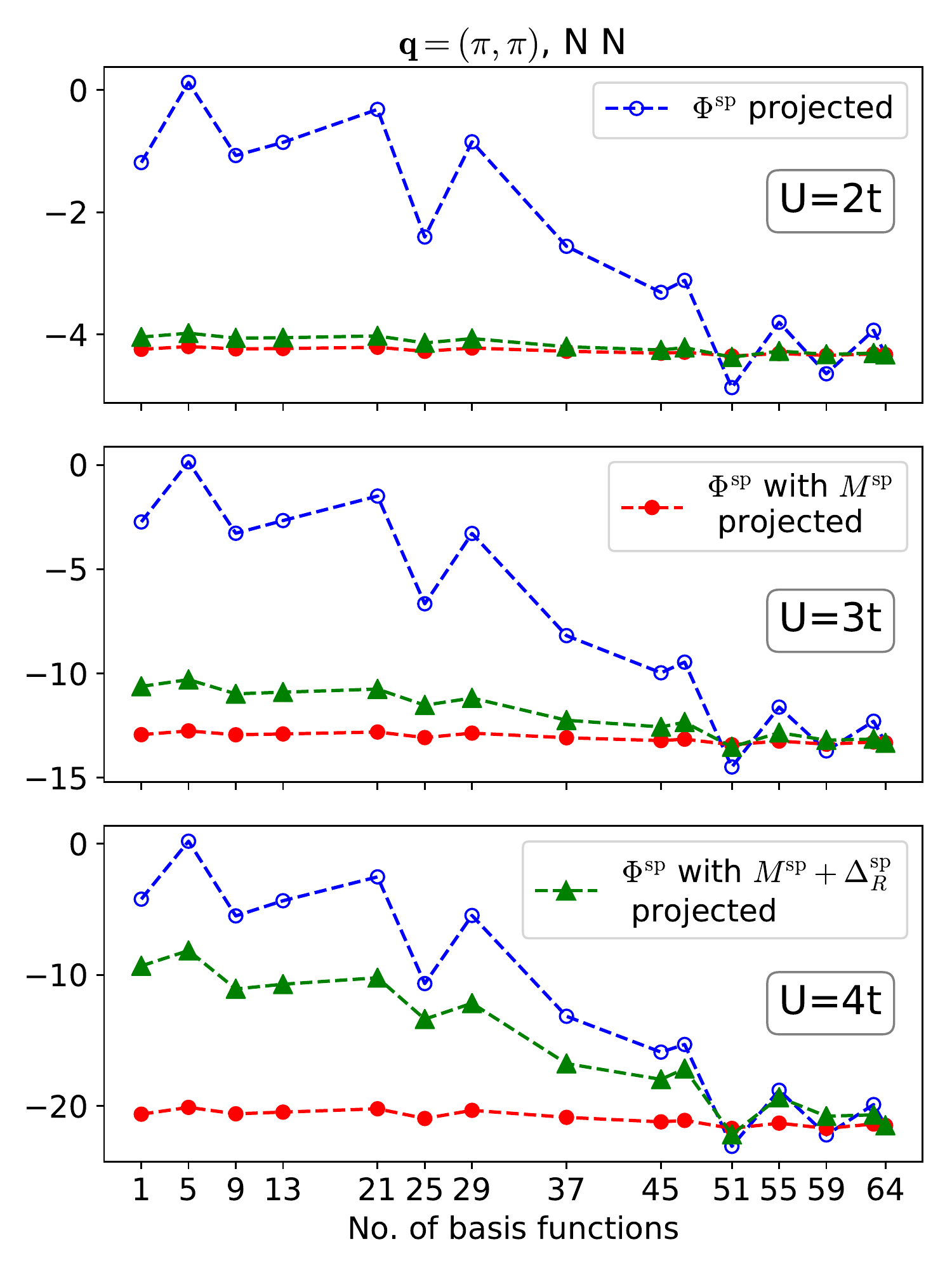}};
    \end{tikzpicture}
\end{center}
    \caption{\label{fig:ff} Convergence of the truncated unity applied in three different ways:
    Blue lines show the direct application to $\Phi^\sz$, cf. Ref.~\cite{Eckhardt20},
    red lines the application only to $M^\sz$, cf. Refs.~\cite{Krien20,Krien20-2}.
    Green lines indicate application to $M^\sz+\Delta^\sz_R$,
    the rest function of the vertex asymptote~\cite{Wentzell20}.
    }
\end{figure}

To this end, we expand the $\qv$-dependence of various vertices in the form-factor basis~\cite{Eckhardt20}
and observe the convergence with respect to the number of expansion coefficients;
see also Ref.~\cite{Krien21-2} where this was done for $U/t=2$ and $T/t=0.2$.
To keep the maximal number of form factors $f(\ell,\qv)$ small we use results for an $8\times8$ lattice.
We transform, for example, $\Phi^\sz$ to the form-factor basis and back into $\qv$-space,
after discarding all but $N_\ell$ form factors,
\begin{align}
\Phi^\sz(\qv,N_\ell)\equiv
\sum_{\ell=1}^{N_\ell}f^*(\ell,\qv)\sum_{\qv'} f(\ell,\qv')\Phi^\sz(\qv'),
\end{align}
where we set $\nu=\nu'=\pi T, \omega=0, \kv=\kv'=(\frac{\pi}{2},\frac{\pi}{2})$ fixed.
The complete $\qv$ dependence is thus recovered for $N_\ell=64$,
but the series may be truncated at a smaller $N_\ell$ if the
expanded vertex is sufficiently short-ranged in real space (truncated unity).
Blue lines in Fig.~\ref{fig:ff} show for $\qv=(\pi,\pi)$ the thus expanded $\Phi^\sz$,
the reducible vertex of the traditional parquet formalism, for $U/t=2,3,4$.
Notice that in the considered regime the antiferromagnetic
correlation length $\xi$ increases monotonously with $U$.
{Namely, we find for $U/t=2$ and $3$ that $\xi\approx1.5$ and $2$, respectively,
which are consistent with our calculations for the $16\times16$ lattice.
For $U/t=4$ we expect a sizable finite-size effect for the $8\times8$ lattice~\cite{Klett20},
which arises for $\xi$ on the order of half the linear lattice size or larger.}

Since the form-factor expansion of $\Phi^\sz$ with respect to $\qv$ converges only slowly,
Ref.~\cite{Krien20} introduced the idea, within the bosonized parquet approach,
to expand only the multi-boson exchange $M^\sz$ in form factors
while the full momentum dependence of $\Delta^\sz$ was kept.
Using Eq.~\eqref{eq:phi} this corresponds to the approximation
$\Phi^\sz(\qv)\approx M^\sz(\qv,N_\ell)+\Delta^\sz(\qv)-U^\sz$.
Red lines in Fig.~\ref{fig:ff} show this result again for $\qv=(\pi,\pi)$.
For $U/t=2,3,4$ this approximation lies close to the fully converged $\Phi^\sz$ even for $N_\ell=1$.
This indicates, remarkably, that the relative importance of $M^\sz$ compared to $\Delta^\sz$
does not increase with $\xi$ at all
(even if the correlations described by $M^\sz$ grow in range as $\xi$ increases,
they do not grow faster than it is the case for $\Delta^\sz$).

On the other hand, we show now that the relative importance of the term $\Delta_R^\sz$ compared to $\Delta^\sz$
{\sl does} increase with the correlation length.
To this end, we expand this term together with $M^\sz$,
such an approximation may be written as
$\Phi^\sz(\qv)\approx M^\sz(\qv,N_\ell)+\Delta_R^\sz(\qv,N_\ell)+(\Delta^\sz(\qv)-\Delta^\sz_R(\qv))-U^\sz$.
This corresponds to a parametrization of $\Phi^\sz$ where its high-frequency limits are given
through the vertex asymptote, $\Delta^\sz-\Delta^\sz_R$, retaining full momentum dependence,
while the rest function $M^\sz+\Delta_R^\sz$ is expanded in form factors.
The convergence of this parametrization can be observed in the green lines drawn in Fig.~\ref{fig:ff}.
As expected, the convergence with form factors worsens considerably as the correlation length increases
at larger $U/t$, in fact, for $U/t=4$ it becomes comparable to the slow convergence of $\Phi^\sz$.
We conclude that it is advantageous to keep $\Delta_R$ parametrized through $\Delta$,
rather than to combine it with $M^\sz$ in a rest function.

\subsection{Remarks}
Several remarks are in order to put the result reported in Fig.~\ref{fig:ff} into perspective.
Firstly, we recall that the truncated unity is intended foremost to parametrize the
dependence on fermionic momenta $\kv,\kv'$, which is often much weaker than the $\qv$ dependence.
However, an unbiased approach to two-particle correlations, such as parquet or fRG schemes,
requires channel projections which map the $\qv$ dependence in one channel to the $\kv,\kv'$
dependence in another. It was therefore noted in Ref.~\cite{Eckhardt20} that the truncated unity cutoff
unfortunately also appears in bosonic arguments.
This explains the fast convergence of the truncated unity in Refs.~\cite{Krien20,Krien20-2},
where it was only applied to the $M$'s.
In this respect it is also encouraging that the relative importance of $M^\sz$ compared to $\Delta^\sz$
appears to be independent of the correlation length (Fig.~\ref{fig:ff}),
so that the quality of a fixed truncated unity cutoff $N_\ell$ does not deteriorate with growing $\xi$.
Compared to the traditional parquet formalism, the improved performance of our
implementation, and the generally weaker momentum dependence of the quantities calculated in it,
are reminiscent of similar observations in the context of
vertex-corrected $GW$ approaches~\cite{Kotliar06,Kutepov16}.

On the other hand, one has to keep in mind that the practical advantage of the bosonized formalism depends on the physical
regime and the correlation functions of interest.
For example, we find in the half-filled Hubbard model at weak coupling
that $\Delta^\sz$ is much larger than $M^\sz$, however, in the charge channel we find the opposite in the
low-frequency regime. In particular in applications to pseudogaps induced by spin-density wave
fluctuations the charge sector is of a lesser interest, however,
it remains to be seen how much improvement the bosonized formalism offers in other physical settings.
One may hope that in a regime which exhibits strong charge fluctuations the importance of $\Delta^\ch$ may
be enhanced over $M^\ch$.

However, a case where a breakdown of the fast convergence of the truncated unity can be expected is,
for example, a regime of long-ranged $d$-wave singlet fluctuations.
They are captured by the corresponding $M^\sing$ of the particle-particle channel~\cite{Bonetti22}.
How much the results suffer from this may depend on the importance
of the feedback of the $d$-wave fluctuations on other channels,
which requires a projection operation, as discussed above.
In this regard, it is intriguing to consider a re-bosonization and suitable parametrization
(through new $\gamma$'s and $W$'s) of the corresponding strongly fluctuating channel captured by the $M$'s.
As the example of the $d$-wave shows, the bosonized formalism does not come with
an autopilot for improved performance.
However, in any case the interpretative advantages of the bosonization remain,
and there are, to our knowledge, no disadvantages associated with it.

\section{Conclusions}\label{sec:conclusions}
{We applied the parquet approximation to the Hubbard model on a $16\times16$ lattice}
and presented two-particle correlation functions corresponding to the bosonized parquet formalism introduced in Refs.~\cite{Krien20,Krien21-2}.

The vertex functions reveal intriguing patterns as a function of the momenta,
and the few shown examples scratch only
the surface of the diverse variations that we observed in our calculations.
It is an exciting outlook to consider the effects of next-nearest neighbor hopping,
doping, larger interaction~\cite{Krien20,Chalupa21}, and other modifications,
where one or the other of the patterns may emerge as a physically important one.

We applied the truncated unity to quantities defined in the bosonized parquet formalism
and benchmarked its convergence with the number of form factors.
Similar to Ref.~\cite{Harkov21} our analysis reveals that, in the considered setting,
the formalism extends the asymptotic parametrization of the vertex functions~\cite{Wentzell20}
in a practically useful way to low frequencies.
{In particular, it facilitates fast convergence of the truncated unity approximation
even in presence of long-ranged antiferromagnetic correlations.}

Our implementation can be used to investigate properties of parquet-based approximations in their pure form
for reasonably large lattice sizes, such as the fulfillment of Ward identities~\cite{Janis17,Chalupa21-2}
or nontrivial sum rules for the vertex functions~\cite{Mermin67},
without any additional approximations.
\\
\\
\noindent
\textbf{Acknowledgments}\
We acknowledge financial support from the Austrian Science Fund (FWF) through Projects No. P32044 and No. P30997.

\section*{Author contributions}
A.K. implemented the algorithm of Ref.~\cite{Krien21-2}. Both authors analyzed the results and jointly prepared the text.

\bibliography{main}
\bibliographystyle{epj}

\end{document}